# 3D deep learning for enhanced atom probe tomography analysis of nanoscale microstructures

Jiwei Yu [1], Zhangwei Wang [1*], Aparna Saksena [2], Shaolou Wei [2], Ye Wei [3], Timoteo Colnaghi [4], Andreas Marek [4], Markus Rampp [4], Min Song [1*], Baptiste Gault [2, 5], Yue Li [2*]

1. State Key Laboratory of Powder Metallurgy, Central South University, Changsha, 410083, China
2. Max-Planck-Institut für Eisenforschung GmbH, Max-Planck-Straße 1, Düsseldorf, 40237, Germany
3. Ecole Polytechnique Fédérale de Lausanne, School of Engineering, Rte Cantonale, 1015 Lausanne, Switzerland
4. Max Planck Computing and Data Facility, Gießenbachstraße 2, Garching, 85748, Germany
5. Department of Materials, Imperial College, South Kensington, London SW7 2AZ, UK

*Corresponding authors: yue.li@mpie.de (Y. L.); z.wang@csu.edu.cn (Z. W.); msong@csu.edu.cn (M. S.)

**Abstract:** Quantitative analysis of microstructural features on the nanoscale, including precipitates, local chemical orderings (LCOs) or structural defects (e.g. stacking faults) plays a pivotal role in understanding the mechanical and physical responses of engineering materials. Atom probe tomography (APT), known for its exceptional combination of chemical sensitivity and sub-nanometer resolution, primarily identifies microstructures through compositional segregations. However, this fails when there is no significant segregation, as can be the case for LCOs and stacking faults. Here, we introduce a 3D deep learning approach, AtomNet, designed to process APT point cloud data at the single-atom level for nanoscale microstructure extraction, simultaneously considering compositional and structural information. AtomNet is showcased in segmenting $L1_2$-type nanoprecipitates from the matrix in an AlLiMg alloy, irrespective of crystallographic orientations, which outperforms previous methods. AtomNet also allows for 3D imaging of $L1_0$-type LCOs in an AuCu alloy, a challenging task for conventional analysis due to their small size and subtle compositional differences. Finally, we demonstrate the use of AtomNet for revealing 2D stacking faults in a Co-based superalloy, without any defected training data, expanding the capabilities of APT for automated exploration of hidden microstructures. AtomNet pushes the boundaries of APT analysis, and holds promise in establishing precise quantitative microstructure-property relationships across a diverse range of metallic materials.

# 1. Introduction

The overall set of physical properties in materials is governed by microstructures across multiple length scales, spanning from grain-level phase constitution [1-3] down to atomic-level solute distribution [4-6]. Understanding these features necessitates advanced characterization techniques at different length scales for specific applications. Difficulties arise at finer scales, largely owing to the more significant challenges in balancing spatial resolution and statistical reliability [7, 8]. Atom probe tomography (APT), with excellent elemental sensitivity and near-atomic resolution [9], can perform a quantitative 3D assessment of nanoscale microstructures in engineering materials and as such allows for direct correlation with macroscopic properties. This includes not only nanoscale precipitates [10-12], complex oxides [13, 14] and multiphases [15, 16], but also crystalline defects such as dislocations [17, 18] and grain boundaries [19, 20]. Their successful detection generally depends on a certain degree of compositional segregation. Relevant algorithms include isosurface analysis, K nearest neighbor, and radial distribution function to visualize or indicate the degree of segregation [21-24]. An obvious limitation occurs when there are subtle or even negligible elemental segregations, as can be the case for local chemical ordering (LCO) [25] or 2D stacking faults [26] that have been reported as challenging to analyze. Finding ways to exploit the partial structural information within APT data becomes crucial to characterizing those elusive microstructural features.

Following reconstruction, APT data takes the form of a 3D point cloud along with the atomic or ionic identity [27, 28]. APT's chemical sensitivity is in the range of 10–100 ppm, but its spatial information is anisotropic due to the trajectory aberrations (the resolution of the best scenario is 0.3 nm in the lateral direction and 0.1 nm in the depth direction) [29, 30]. In addition, 20–65% of the ions are randomly lost, because of the limited open area of the particle detector or due to grids with limited transparency on the path of the ions. As a result, the mining of the remaining high-quality yet partial crystallographic information requires expertise, sophisticated tools, and remains both time-consuming and user-dependent [31-34]. Machine-learning-based algorithms are being developed to improve data extraction, simplify and automate data analyses, including user-independent mass spectrometry analysis [35-38], intelligent interface detection [19, 39, 40], or more complete APT data analysis workflow [41, 42].

Another example is the analysis of the partial structural information retrained within APT data [43]. This is typically analyzed through the use of spatial distribution maps (SDM). The generation of a single SDM requires the integration of signals from a certain volume of atoms (1–2 nm), and the 3D structural data is reduced to 1D- of 2D-histograms enabling the quantification, along specific crystallographic orientations, of inter-atomic distances. Systematically generating and analyzing SDMs large datasets from which millions of SDM patterns require automated analysis workflows as recently enabled by machine-learning [44, 45], including for challenging LCO detection [25, 46, 47].

With APT's point cloud being intrinsically 3D, it is natural to extend these methodologies for extracting microstructural features directly to 3D, going beyond conventional analyses that start with a data dimensionality reduction into 1D/2D descriptors, causing potential information loss [25, 31]. Our recent work demonstrated the possibility of applying a 3D convolutional neural network (CNN) to analyze voxelized APT data to segment the 3D distribution of $L1_2$-type nanoprecipitates from a disordered FCC matrix [48]. However, the required data region was limited to specific crystallographic poles and did not address situations where no clear elemental segregations exist. Moreover, the nature of voxelization limited the size range of the recognized domain using CNN, thus not reaching down to the single-atom scale.

Here, we propose a 3D point-cloud-based neural network, named AtomNet, to handle the information at the single-atom level without voxelization to reveal different nanoscale microstructures. AtomNet is based on PointNet [49], which can effectively and robustly handle point cloud data. Prior knowledge about continuous phase distribution is also introduced to AtomNet for better recognition ability. First, AtomNet is tested by 3D imaging nanoprecipitates in an AlLiMg alloy used as a benchmark, and we showcase its ability beyond previous work that would only work in the highest spatial resolution. Then, the more challenging case of the detection of $L1_0$-type LCOs in red gold, which has a composition close to equiatomic AuCu, for which previous segregation-based analysis isosurface failed to identify ordered domains. Finally, the ability of AtomNet to indicate the positions of stacking faults is explored in a Co-based superalloy. The advantages and limitations of AtomNet are discussed, along with directions for future developments.

## 2. Materials and methods
### 2.1. Materials

APT data of Al-6.79Li-5.18Mg (at.%, thereafter) alloy annealed at 150°C for 8h was selected as a benchmark, as it has been previously used in Ref. [31, 48, 50]. The data was collected on the Cameca LEAP 3000XSi with a 55% detection efficiency [50]. A deformed Au-46.8Cu-5.3Ag red gold [51] and Co-based superalloy (Co-32Ni-8Al-5.7W-6Cr-1.8Ta-2.8Ti-0.1Hf-0.4Si) alloy were chosen to show the LCO and defects recognition ability of AtomNet, respectively. The former APT measurement was performed on a LEAP 5000XS with an 80% detection efficiency, while the latter was on a LEAP 5000XR with a 52% detection efficiency. All site-specific (along the {002}) needle-like specimens were prepared using the FEI Helios focused ion beam with a Ga ion source. The APT experiments were performed in laser pulsing mode at 50-60 K, 0.8-1.0 % detection rate, 40-45 pJ laser energy, and 125-250 kHz pulse rate. APSuite 6.3 was used for all initial reconstructions by tuning two important parameters, i.e. the field factor and image compression factor according to the method introduced in Ref. [52, 53].

### 2.2. Feature engineering

Appropriate feature engineering is a cornerstone of machine learning [54]. APT data has two primary components: the Euclidean spatial coordinates ($X$, $Y$, and $Z$) of each atom of the point cloud; the other is the mass-to-charge information of each atom to identify the chemical species. Inspired by how scientists distinguish different crystal structures with specific elemental site occupations, a simple and efficient feature extraction method is proposed. For each atom, we extracted its relative 3D atomic position relative to the nearest neighbor (NN), i.e. $\Delta X$ $\Delta Y$ $\Delta Z$, as shown in Fig. 1a. *N* (neighbor) and *S* (self) represent the elemental species of neighboring atoms and the selected atom itself, respectively. Here we use A, B, C … to represent different elements. After careful tuning, it was found that the relative coordinate (*X*, *Y*, and *Z*) and compositional information (*N* and *S*) from the 32NNs of each atom, which falls between the second layer (18NNs) and third layer (42NNs) of FCC structure, are appropriate for the following training procedure. Ultimately, an input feature with the shape of (32, 5) was extracted for each atom.

**2.3. Workflow of AtomNet**

PointNet is a popular 3D neural network that handles point clouds directly and efficiently, respecting the permutation invariance of inputs [49]. This avoids generating large amounts of sparse data and losing information that was encountered when using the CNN-based strategy to transform point clouds into regular 3D voxels, 2D images or 1D curves [31, 48]. Here, we propose AtomNet, which utilizes PointNet as a fundamental building block, to identify challenging nanoscale microstructures in APT data. As shown in Fig. 1b, the applied PointNet block mainly consists of 2 T-Nets, 3 MLPs and 1 MaxPool layer. The T-Net allows affine transformations such as translation, rotation, shearing, and so on. It was originally designed to ensure 3D spatial invariance. T-Net consists of 3 Conv1D, 2 Dense, 1 MaxPool and 1 Transformation layers, followed by the original input dotting with a transformation matrix to complete the affine transformation. MLP stands for the multi-layer perceptron with 2 to 3 weights shared Conv1D or Dense layers. MLP is the primary computational processing unit, and traditional artificial neural networks could be constructed only with it. MaxPool is a pooling or aggregating layer that stores the maximum value and discards others, providing interactions within nearby atoms. After training the first PointNet block, AtomNet shows moderate predictive ability, for example, with the obtained AUC (area under the receiver operating characteristic curve) value being 0.78 in the Au-Cu alloy.

To further improve model performance, a feature updating strategy was utilized to introduce prior knowledge, as an inductive bias [55] to help model learn specific notions efficiently. Here, the notion refers that precipitates, LCOs or distinct phases usually consist of continuous and compact groups of atoms. If most of the neighbor atoms belong to the specific phase, the target atom would also belong to the same phase. The original features were updated after the first PointNet block, as shown in Fig. 1b. Then the updated features were fed into the second PointNet block, and a higher AUC of 0.86 was obtained in the Au-Cu alloy. Fig. 1c visualizes the results before and after the first feature update, suggesting that the number of incorrectly predicted atoms has been greatly reduced. Fig. 1d shows the detail of feature update, by adding/updating predictions from the last PointNet block. Theoretically, this iteration can be repeated to get better performance, at the expense of heavy computations. A detailed test is provided in Section 2.5.

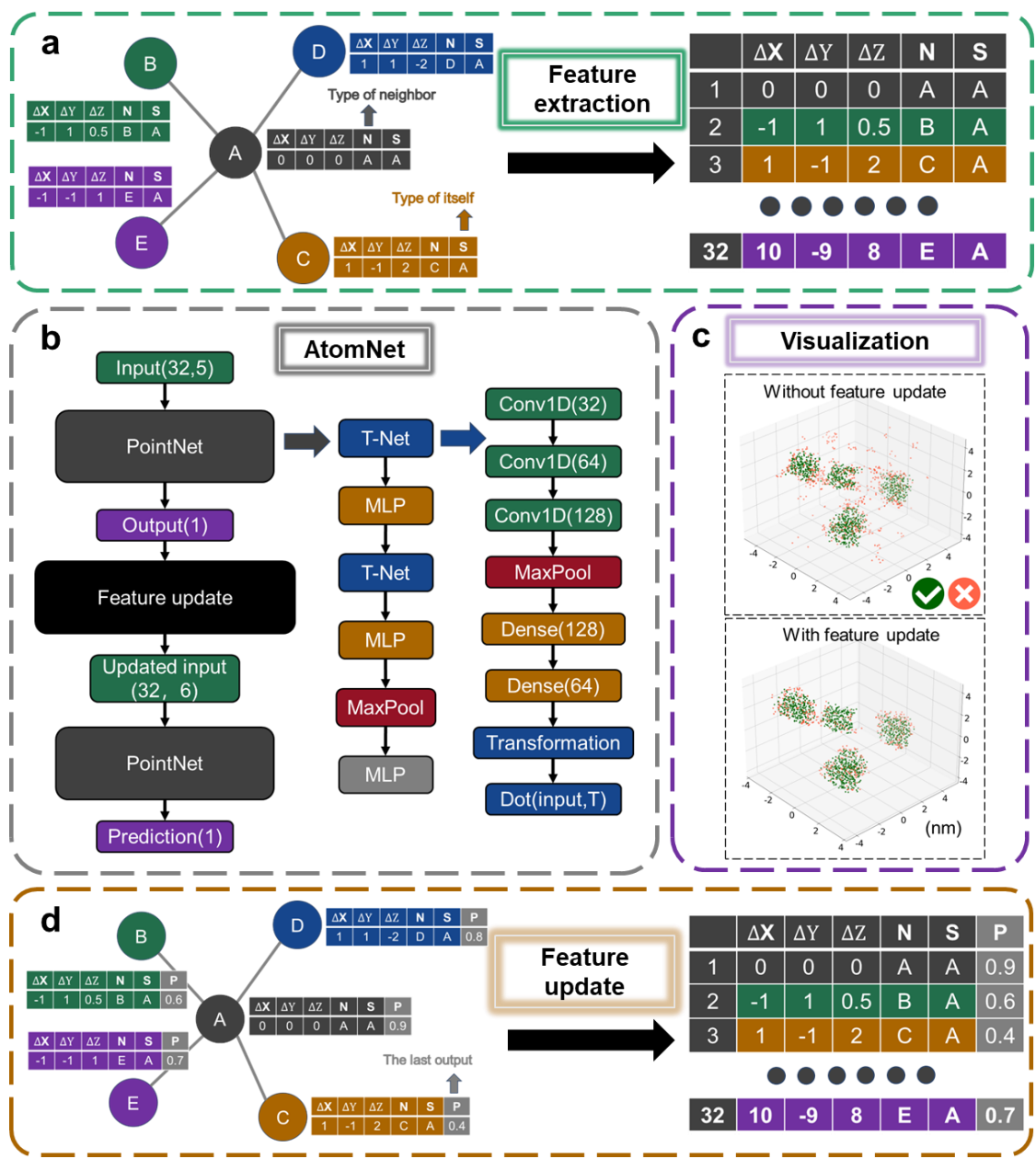

**Fig. 1** Overview of the proposed AtomNet architecture. (a) The details of feature engineering of each atom. The left part shows the schematic of sampling information from neighboring atoms in APT data. The right part shows the input feature matrix of the "A" atom. (b) Architecture of AtomNet model. From left to right, AtomNet, PointNet block and T-Net are painted successively. Brackets in AtomNet indicate data shape and in T-Net mean the number of filters/neurons. All activation functions are "Relu" except the output layer which uses "Sigmoid". (c) Visualization of nanoparticles before/after feature updating. Green atoms represent a correct prediction, while red atoms represent an incorrect prediction. (d) Feature updating for incorporating prior knowledge to improve accuracy.

### 2.4. Simulated data bank

As a supervised algorithm, AtomNet requires reliable training datasets. Here, a pipeline is proposed to generate synthetic APT point cloud datasets by simulating the
5

trajectory aberration and imperfect detection efficiency encountered in APT experiments [52]. As shown in Fig. 2, a perfect 3D FCC-matrix superlattice was built with a size of 4×4×4 nm$^3$ at first. Then, different sizes of spherical L1$_2$ (Al$_{0.75}$Li$_{0.2}$Mg$_{0.05}$), L1$_0$ (AuCu), and L1$_2$ (Co$_{0.4}$Ni$_{0.35}$Al$_{0.095}$W$_{0.043}$X, X representing all remaining elements) nano-domains were embedded into an FCC matrix of Al-Li-Mg, Au-Cu and Co-based superalloys, respectively. The detailed size information for each system is listed in Table 1. The matrix composition was based on APT experimental data, while the secondary phase composition was approximated by rounding the elemental ratio to 3:1 for L1$_2$ and 1:1 for L1$_0$. Note that the compositions of L1$_2$ (Al$_{0.75}$Li$_{0.2}$Mg$_{0.05}$) and L1$_2$ (Co$_{0.4}$Ni$_{0.35}$Al$_{0.095}$W$_{0.043}$X) were determined based on the measured APT data. Third, the superlattice was rotated to simulate experimental data along different crystallographic orientations, allowing to simulate regions of high and low spatial resolution (i.e. near and away from crystallographic poles). As listed in Table 1, AlLiMg alloy was randomly rotated to ±90° to simulate all possibilities of crystallographic orientations. While for the other two systems, the rotation along the x and y axis was limited to ±3° and along the z axis was randomly chosen between ±90° to simulate experimental data along the {002} pole. Note that the local rotations along x and y are intended to reflect the local distortion of the atomic planes along the pole [56]. Finally, atoms were shifted in 3D according to different degrees of Gaussian noise and a certain fraction of atoms were removed randomly to simulate the anisotropic spatial resolutions and imperfect detection efficiency (See Table 1).

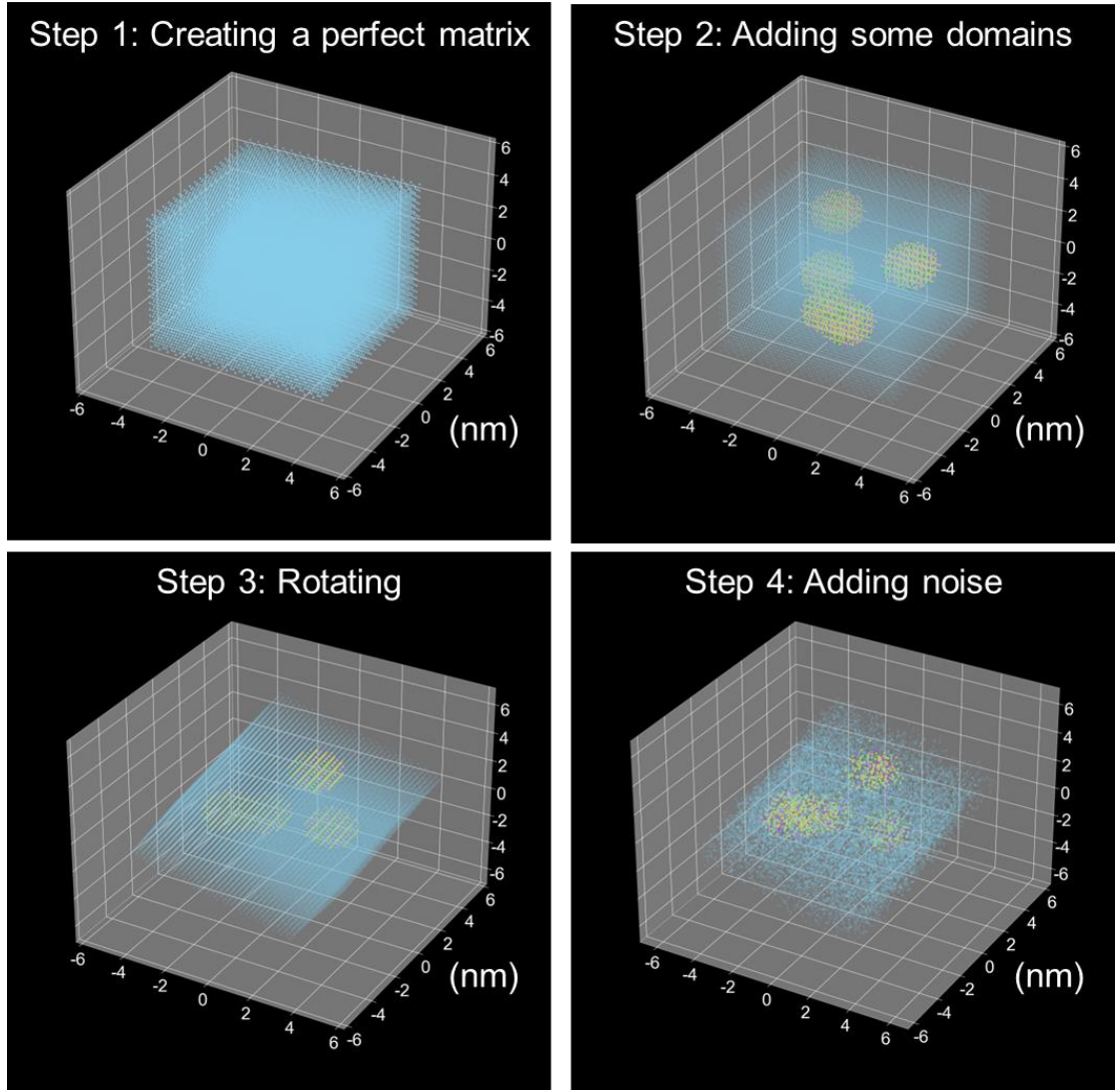

**Fig. 2** Pipeline of simulated APT data. Adding some specific nano-structures ($Al_{0.75}Li_{0.2}Mg_{0.05}$ nanoparticles here) into a perfect FCC matrix, then rotating and disturbing to simulate the crystallographic orientation and trajectory abbreviation encountered in APT data, respectively.

Table 1 Parameters of simulated data in different alloying systems.

| Alloys | $L1_2/L1_0$-domain radius (nm) | Rotation (°) | | Detection efficiency | Trajectory aberration (nm) | |
|---|---|---|---|---|---|---|
| | | X/Y | Z | | X/Y | Z |
| **Al-Li-Mg** | 1.2-2.0 | ±90 | ±90 | 40%-80% | 0.2-0.5 | 0.08-0.02 |
| **Au-Cu** | 0.8-1.2 | ±3 | ±90 | 40%-80% | 0.2-0.5 | 0.08-0.02 |
| **Superalloy** | 1.4-2.0 | ±3 | ±90 | 40%-80% | 0.2-0.5 | 0.08-0.02 |

## 2.5. Training details

AtomNet was implemented by the TensorFlow-GPU 2.10.0 backend on Python 3.9.7. Training/validation/testing data contained 80/10/10 cubes with a 4/4/10 nm length, respectively. Each 4-nm cube included approx. 3000 atoms. After a thorough tuning procedure, the chosen loss function was 'BinaryCrossEntropy' and the optimizer was 'Adam' with a learning rate of $10^{-3}$. AUC was used as the metric to measure unbalanced datasets (the ratio of atoms between the labeled $L1_2/L1_0$-domain and the matrix is close to 1:4). For the training procedure, the chosen batch size was 256, and callbacks were used to monitor and save the best model. An Au-Cu example of the evolution of AUC and loss is shown in Fig. 3. The change in background color indicates the feature updating by training another PointNet block. The obtained AUC value in the validation dataset increases significantly from about 0.78 to 0.86 after the first update due to the introduced prior knowledge, while the loss increases from about 9.22 to 9.45 at the same time. The higher loss may be related to the newly added dimension, which introduces more variables and uncertainties. Thus, we pay more attention to AUC instead of loss when evaluating the obtained classification model [57]. Based on the balance between the AUC value and computation costs, the number of PointNet blocks in the Al-Li-Mg, Au-Cu, and Co-based superalloy is 2, 3, and 2, respectively.

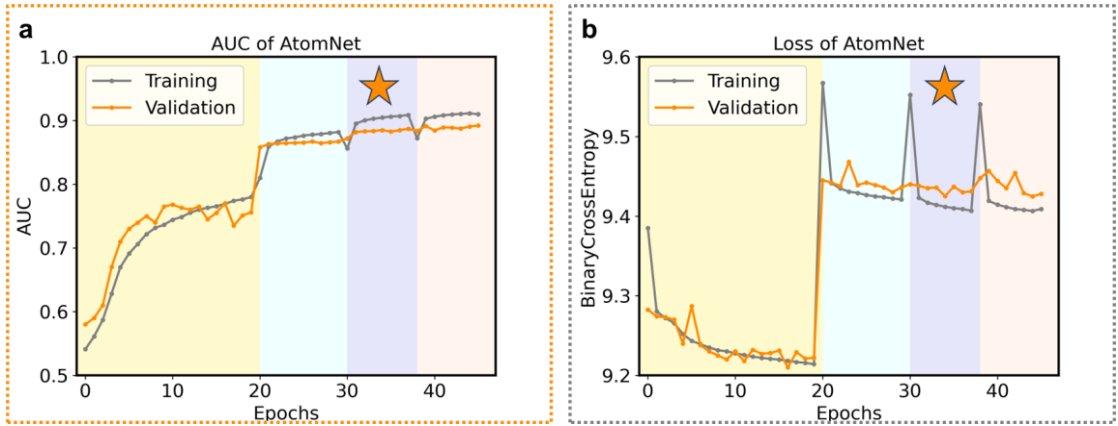

**Fig. 3** Training and validation of AtomNet. (a) AUC and (b) loss of AtomNet with epochs in the Au-Cu system. Loss/BinaryCrossEntropy is a differentiable function that helps gradient descent during training and judges whether the overfitting occurs. Metric/AUC is the decisive factor of model performance. The background color will change after adding another PointNet block with the updated features. The orange star marks the final choice.

## 3. Application of AtomNet
## 3.1. Nanoprecipitates in AlLiMg

Nanoscale precipitates play a critical role in influencing the mechanical properties of alloys primarily through precipitation hardening mechanism [1, 2, 12, 58]. Quantifying these nanoprecipitates is beneficial for establishing microstructure-property relationships that can help further in designing advanced materials [59, 60]. The Al-Li-Mg dataset used in Ref. [31, 50] was used as a benchmark to test AtomNet.

Fig. 4a shows an example of the simulated L1$_2$-type Al$_3$(Li, Mg) particles with a radius of 1.2-2 nm embedded in a disordered FCC matrix. AtomNet accurately predicted if an atom belongs to the L1$_2$ phase, as shown in Fig. 4b. After performing once feature updating strategy mentioned in Section 2.3, the final AUC score is $0.890 \pm 0.033$. The AUC reflects how well the model is trained on unbalanced datasets (See Section 2.5), but not sufficient to judge the recognition ability for each class.

Recall and precision metrics were used to further assess the recognition ability, as displayed in Fig. 4c and d. Recall is a metric that reflects how many positive samples can be detected, while precision tracks the reliability of predictions, defined as below:

$$\text{Recall} = \frac{\text{Total number of correctly predicted oredered atoms}}{\text{Total number of truely oredered atoms}} \quad (1)$$

$$\text{Precision} = \frac{\text{Total number of correctly predicted oredered atoms}}{\text{Total number of predicted oredered atoms}} \quad (2)$$

AtomNet obtained an overall recall and precision of 0.79 and 0.72 (Fig. 4), respectively. It is interesting to note that both recall and precision of Li are higher than those of Al and Mg elements, which can be explained by its higher tendency for partitioning between the ordered precipitates (20 at.%) and the disordered matrix (5 at.%). This first example showcases AtomNet's capability for reliably identifying L1$_2$-ordered precipitates in simulated datasets.

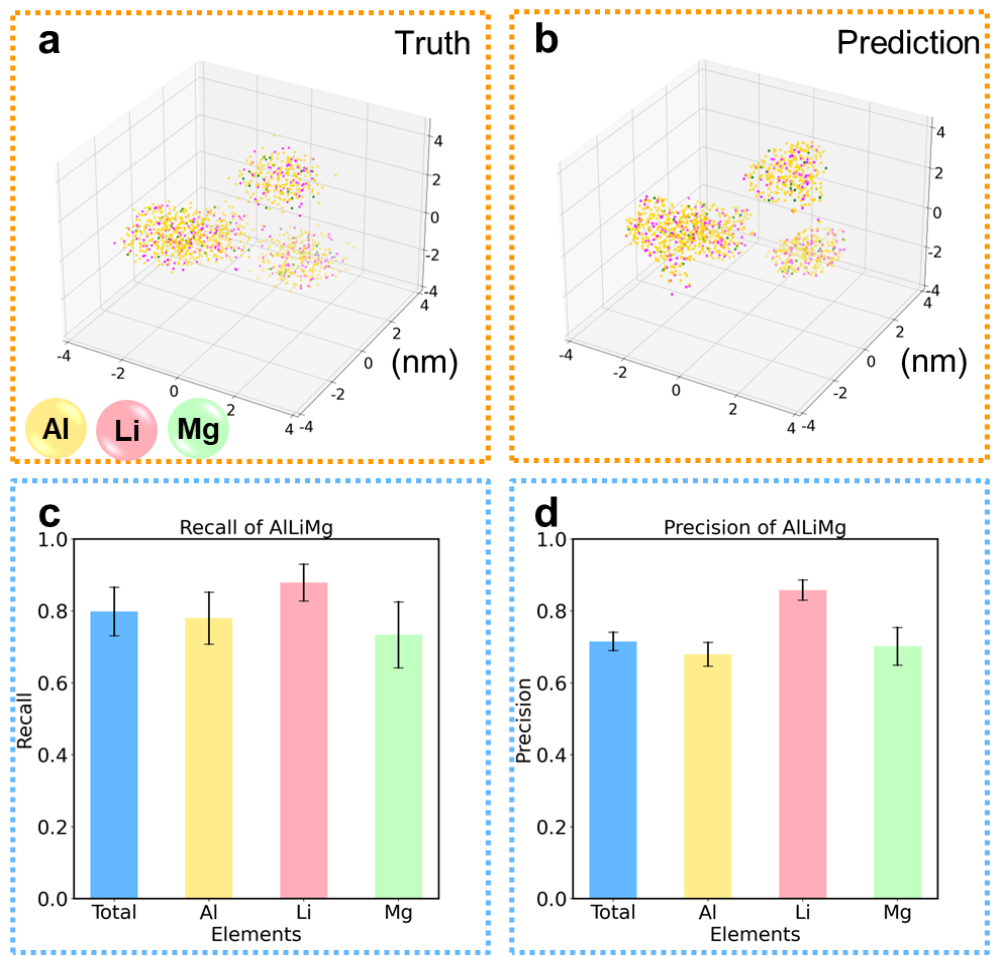

**Fig. 4** Performance of AtomNet on the simulated AlLiMg test datasets. (a) An example of some simulated $L1_2$ nanoprecipitates and (b) corresponding predicted results. Only atoms from nanoprecipitates are shown. (c) Recall and (d) precision from different elemental species, respectively. 10 cubes with a length of 10 nm were analyzed to obtain a statistical result with the default classification threshold of 0.5 (a more detailed discussion about the choice of thresholds will be given later).

AtomNet was then applied to the experimental data, both close to and away from the {110} pole, in regions-of-interest indicated in the detector hit map shown in Fig. 5a. Analysis of structural information in APT can normally be done only in regions near poles, i.e. where the corresponding crystallographic planes are imaged. Fig. 5b shows nanoprecipitates captured by AtomNet, at the {110} pole. Fig. 5c shows the distribution of $L1_2$ particles marked by an 8 at.% Li isosurface. The AtomNet prediction closely matches the isosurface result, with similar average size and spatial locations.

AtomNet is also consistent with the previous CNN method (Fig. 5d) [31]. It's a voxelization method and can only work along specific pole sites. The prediction is based on cubes of a certain size, so the interface between the matrix and the nanoparticles is facetted. Instead, AtomNet sets out from every single atom and thus can retain the intrinsic nature (nearly atomic resolution) of APT data. Moreover, the blue box in Fig. 5d marks one missing nanoprecipitate via CNN, although it ought to exist (compared with Fig. 5b and Fig. 5c). Third, as compared to previous CNN result, some noisy points/atoms remain via AtomNet due to its single-atom nature, which may be eliminated with a clustering algorithm by setting the minimum number of atoms in a cluster.

We further obtained the spatial distribution maps along the depth direction (z-SDMs) [61] of Al-Al pairs of the recognized $L1_2$ particles and remaining matrix, as shown in Fig. 5e. The matrix has an interplanar distance of 0.14 nm while the $L1_2$ nanoparticle has an interplanar distance of 0.28 nm. This difference relates to the atomic occupancy of Al, which is random in FCC and face-centered in $L1_2$, as discussed in Ref. [50].

Poles are not always visible or planes imaged, and this has limited the application of previous approaches to small reconstructed volumes near poles. As Fig. 5f and g reveal, AtomNet can also detect precipitates in the subset of the data in which atomic planes are not imaged, and these agree well with segmentation based on isosurface. Although AtomNet was trained using spherical particles, this example demonstrates that non-spherical domains (Fig. 5f) can be identified.

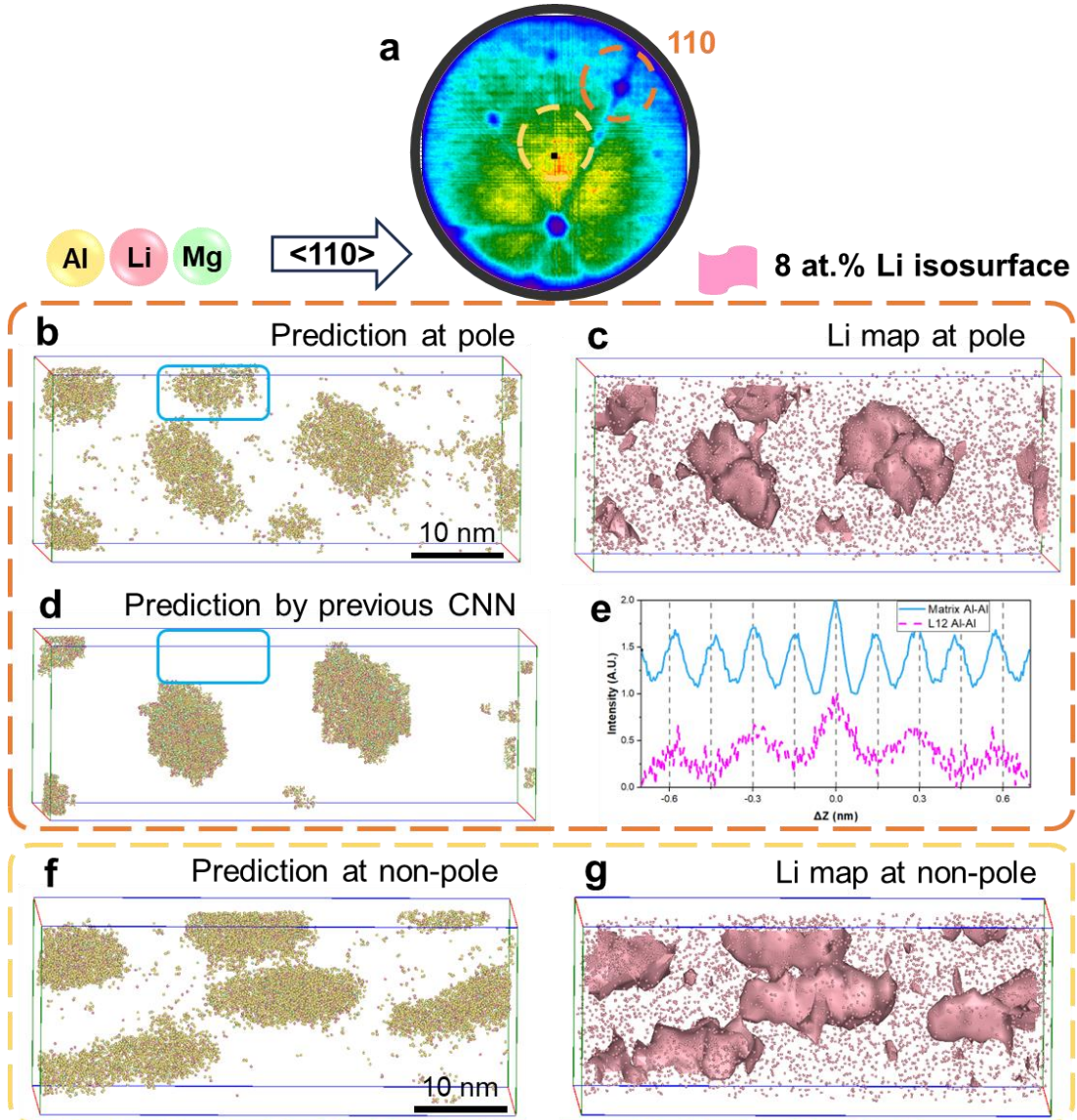

**Fig. 5** Validations on AlLiMg experimental data. (a) 2D detector hit map to highlight the pole and non-pole sites for further analysis. The orange circle marks {110} pole, while the yellow circle is non-pole. Reconstructions are based on pole (b, c, d) and non-pole (f, g) respectively. (b) is the prediction by AtomNet. Here only nanoprecipitates are displayed. (c) is the isosurface method to show nanoparticles on Li maps. The concentration threshold is 8 at.% Li. (d) Prediction with previous CNN method [31]. Blue box in (d) marks a missing nanoprecipitate. (e) z-SDMs of Al-Al pair of nanoprecipitates and the remaining matrix from (d). (f) and (g) are predictions by AtomNet and isosurface respectively, at non-pole sites.

### 3.2. Local chemical orderings in red gold

The last case study exhibits obvious elemental segregations, which can also be handled using other approaches. Here, we will further explore the performance of AtomNet in a more challenging case with LCOs. Red gold is generally considered to have a transformation from FCC to the ordered $L1_0$ phase, which hardens the material but limits its workability [51]. An interesting shape memory effect is also related to this $L1_0$ phase [62]. In-situ synchrotron X-ray diffraction technique indicated the presence of LCOs by observing weak peaks [51] or peak dissymmetry [63], however the characterization of the early stage of $L1_0$ ordering remains challenging due to the lack of obvious segregation, coherent interfaces with the matrix, and the small size of the ordered domains (typically below 2 nm). Here we apply AtomNet to capture these tiny $L1_0$-type LCOs. Note that the isosurface approach cannot work here due to the absence of obvious compositional differences between LCOs and FCC matrix.

    First, AtomNet was tested on simulated datasets. As displayed in Fig. 6a, we built several $L1_0$ LCO domains with a diameter of 1.6 nm embedded in FCC matrix. After simulating the 40% detection efficiency, a domain would only contain in the range of 50 atoms, making its detection arduous compared to larger precipitates as in Al-Li-Mg. This analysis achieved an AUC score of $0.822\pm0.052$ on test datasets when using the default threshold of 0.5, while the predicted LCOs experienced significant size shrinkage in Fig. 6b (the average radius is reduced from about 0.8 nm to about 0.6 nm). To improve the shape and size accuracy of predicted LCOs, different thresholds were compared, and finally 0.3 was chosen, as shown in Fig. 6c. A higher threshold of 0.7, Fig. 6d, only leaves a few atoms per domain due to the strict classification criterion. The obtained recall and precision showed a trade-off in Fig. 6e and Fig. 6f. As the threshold increases, the precision tends to increase while the recall tends to decrease. For these difficult tasks, keeping a high recall can ensure that the targeted nanodomains are recognized more completely. A low threshold of 0.3 is preferred in this case.

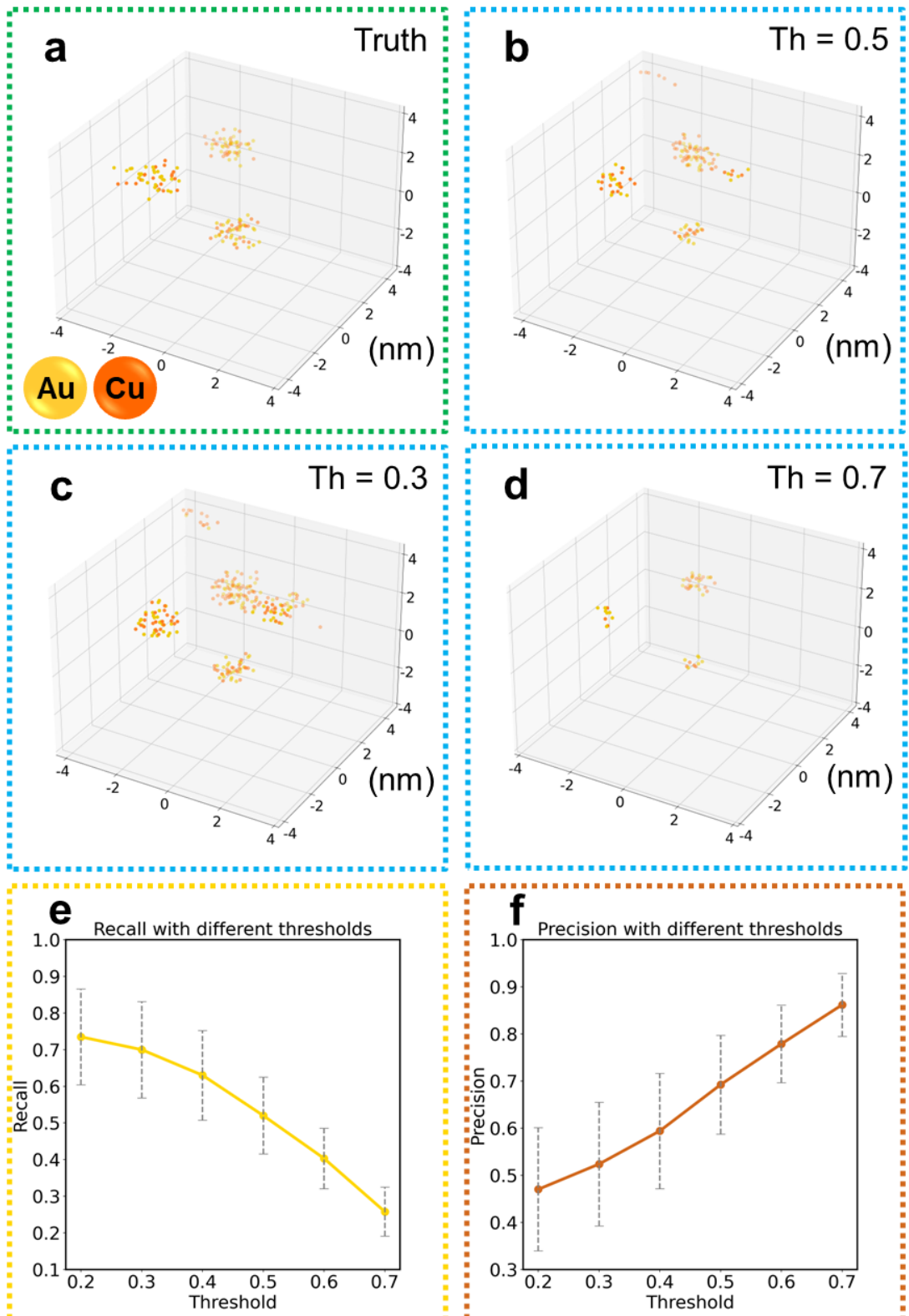

**Fig. 6** Trade-off between recall and precision in simulated Au-Cu test dataset with $L1_0$ LCOs. (a) Ground truth of simulated $L1_0$ LCOs with a diameter of 1.6 nm. The matrix is hidden. (b), (c), and (d) are corresponding predictions by AtomNet with different thresholds (represented by "Th" with the value from 0 to 1). (e) and (f) The evolutions of recall and precision with varied thresholds, respectively. 10 cubes with a length of 10 nm were analyzed to obtain the statistical result.

Fig. 7a shows the obtained distributions of 3D $L1_0$ LCO domains along the {100} pole, and in Fig. 7b each identified domain is displayed with a separate color. Note that the clustering method in APSuite 6.3 was applied to assess the size distribution of LCO domains with a maximum separation distance of 0.4 nm and a minimum number of ions in the cluster of 3. The domain appears spherical. Fig. 7c plots the z-SDMs of different elemental pairs in the matrix (FCC) and LCO domains. Both peak-to-peak distances of Au-Au and Cu-Cu z-SDMs are half than those in LCOs, which is consistent with the expected crystal structures (FCC and $L1_0$). The distribution of size (number of atoms) versus count of LCOs is given in Fig. 7d, and compared with that from a chemically randomized dataset. The latter was generated by maintaining the x, y, and z coordinates but randomly shuffling the mass-to-charge and the associated elemental identities [25, 46]. We compare the size distributions of four randomized datasets with each other based on the contingency coefficient ($\mu$) [46]. The upper limit of the obtained $\mu$ being near 0.25 is regarded as a baseline for these randomized size distributions in experimental data. After analyzing the experimental data, an average value of $0.270\pm0.011$ was obtained, which suggests the occurrence of non-statistical $L1_0$-LCOs in this system. A fraction of LCO domains with a size above 1 nm diameter (>55 atoms) exist in the experimental data while no obvious sign in the randomized dataset.

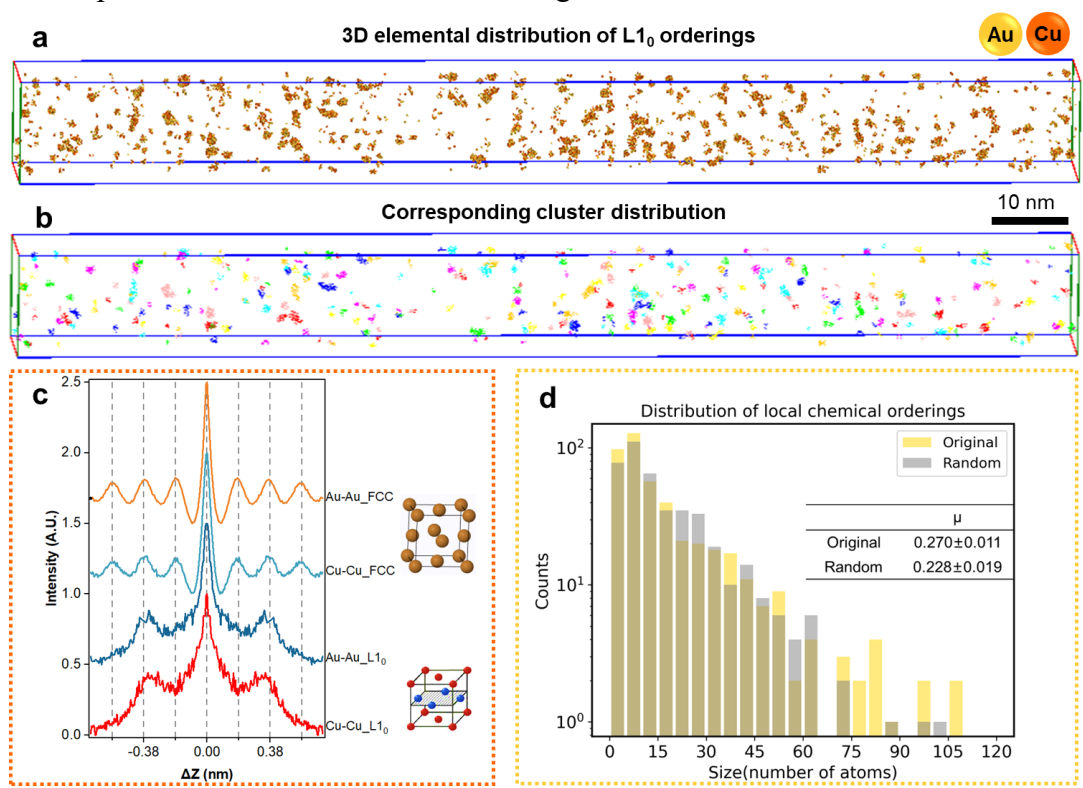

**Fig. 7** 3D distribution of $L1_0$-typed LCOs in Au-Cu along the {100} pole. (a) Elemental distributions predicted by AtomNet. (b) Corresponding cluster distribution in (a). (c) Au-Au and Cu-Cu z-SDMs from $L1_0$-typed LCOs in (a) and remaining FCC matrix. The right part displays the ideal crystal structure of matrix (FCC) and LCOs ($L1_0$). (d) Size distribution of $L1_0$-typed LCOs in (b). A chemical-randomized dataset is compared. The inserted table gives a $\mu$ value, a parameter that indicates the degree of randomization.

### 3.3. Stacking faults in a deformed Co-based superalloy

Stacking faults (SFs) are 2D crystallographic defects along which the stacking of {111} close-packed planes is out of order [64, 65]. SFs reduce the dislocation mobility in single crystal superalloys, and affect their creep responses [66, 67]. Recent studies also suggested that deformation faults could be used to further design metastable alloys [8, 64, 68], as they can serve as the loci for local phase formation and transformation [69-71]. This motivates the third application of AtomNet. Assessing the potential presence of SFs within APT data is however challenging, and has often required the use of correlative electron microscopy [26], and depends on the degree of elemental segregations to the SFs, which is unknown beforehand and can be subtle. Here, we considered a deformed Co-based superalloy as an example to explore the potential of AtomNet on recognizing hidden patterns associated with defects like SFs. The training data only consists of the $L1_2$ phase ($Co_{0.4}Ni_{0.35}Al_{0.095}W_{0.043}X$) and FCC matrix. To ensure AtomNet focuses more on the structural information rather than compositional variations, we set the same concentration in the FCC matrix as that in the $L1_2$ phase. Fig. 8a shows data containing primarily a $L1_2$-ordered γ′ precipitate, with only a small volume of the FCC γ–matrix. Based on previous reports [26, 72, 73], SFs have so far been sought by depletion in the Al concentration projection map, as shown in Fig. 8a, in which subtle planar variations are observable.

Here, two zones marked in Fig. 8a were analyzed using AtomNet to automatically search the sites of SFs. In Zone 1 (Fig. 8b), a reduction in the density map of ordered domains is marked by the red arrow, indicating a zone in which the atomic organization is not the $L1_2$ phase and thus possibly a SF. Zone 2 (Fig. 8c) exhibits a vertical wide zone with a low density of $L1_2$-ordered domains, marked by the gray arrow, which was found to go through the entire volume and corresponds to a crystallographic pole. Another zone with a lower density marked by the green arrow can be associated with a SF. AtomNet can hence indicate the position of defects, even without relevant training data.

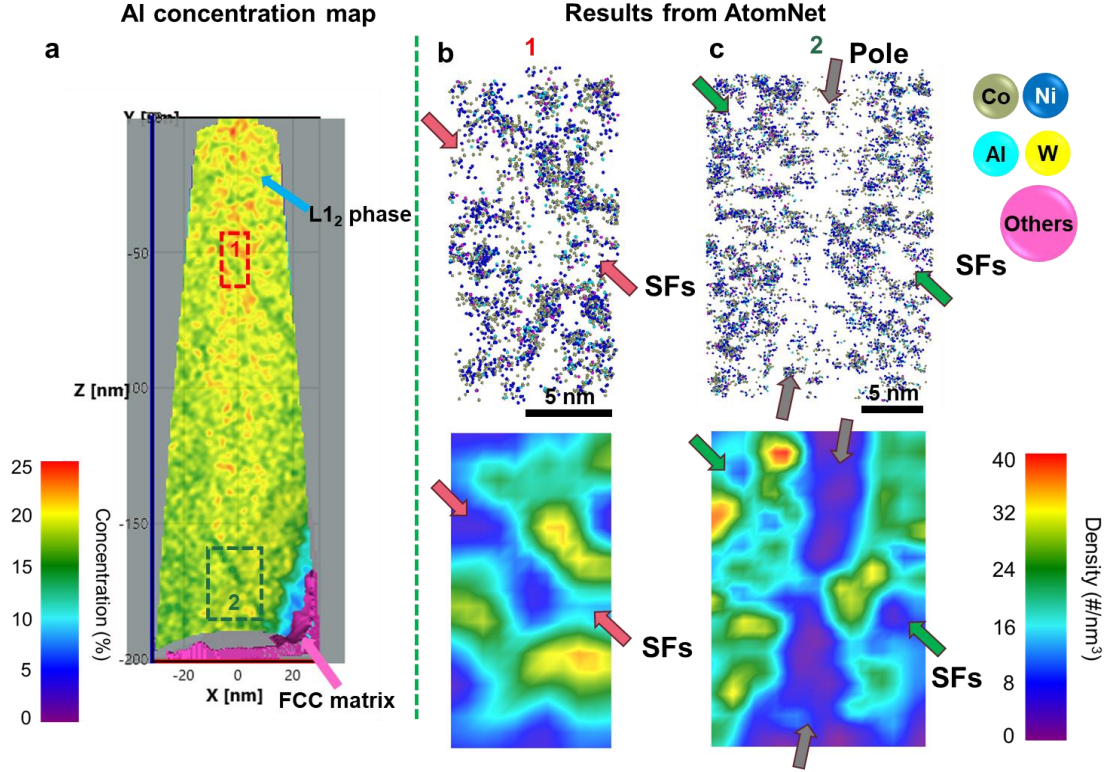

**Fig. 8** Exploration in a deformed Co-based superalloy ($Co_{0.4}Ni_{0.35}Al_{0.095}W_{0.043}X$, X represents remaining elements) with SFs. (a) Traditional 2D Al concentration projection map to indicate potential sites of SFs. Two zones inside the $L1_2$ phase were analyzed by the proposed AtomNet. Predictions of AtomNet in (b) zone 1 and (c) zone 2 with atom (up) and density (bottom) maps. Arrows indicate the sites of SFs and pole. The classification threshold is 0.5.

## 4. Discussion

In this work, a point-cloud-based AtomNet was proposed to intelligently dig out microstructural information hidden within APT data. We successfully applied it in a series of FCC-based case studies with nanostructures spanning from 3D to 2D. AtomNet offers several advantages over previous methods based on isosurface thresholding [27, 28] or CNN-assisted APT analyses [31, 48]. First, unlike previous CNN methods based on voxelization, AtomNet handles every single atom. For instance, the segmentation from AtomNet exhibits a smooth phase boundary while previous methods show an obvious jagged boundary (Fig. 5c, d). Moreover, previous methods partially focus on either the compositional differences, i.e. isosurface, or the structural changes, CNN-APT [31]. AtomNet considers both compositional and structural information simultaneously, by integrating features from 32 nearest neighbors with respect to each atom. These features were transformed and trained via a PointNet block to tackle some challenging situations – including, for instance, precipitates imaged away from regions in which atomic planes are imaged in the case of the AlLiMg alloy (Fig. 5), small LCO domains without obvious compositional segregation in Au-Cu (Fig. 7). Third, AtomNet can indicate nanostructures that do not exist in the training datasets, like SFs in the Co-based superalloy. In this case, the composition of the matrix and precipitates was the same in the simulated data, to focus AtomNet only on the structural information. Thus, AtomNet will respond to those never-seen defects by judging the occupations of atoms. Better performance could be achieved if these defects could be simulated and then used to train AtomNet. Last but not least, AtomNet still works well in non-spherical nanodomains whose composition differs from the simulated one like Figs. 5 and 7, further demonstrating its robustness.

Nevertheless, AtomNet has some limitations. For a specific alloy, relevant training data need to be acquired from either experiments or simulations. While conventional methods like isosurface only require manual analysis and trial and error. Of course, this cannot work in the case of Au-Cu with the existence of LCOs.

For the Au-Cu case, the classification threshold was adjusted to test the performance of the recognition model. A trade-off between the recall and precision (corresponding to the size and count accuracy, respectively) is inevitable (Fig. 6e and f). When choosing a low threshold (Fig. 6c), AtomNet achieves a high recall, i.e., a high size accuracy, which is desired for size-focused research. However, the corresponding precision is low and AtomNet falsely classifies more random atoms as ordered ones, i.e., over-recognition phenomenon. By selecting a high threshold (Fig. 6d), AtomNet will get a high precision, i.e., a high count accuracy, which is better for number-density-focused research. For some tasks like detecting $L1_2$ from FCC in the AlLiMg with obvious segregation and relatively large size (above 2-nm diameter), selecting the default threshold of 0.5 would be appropriate for the most time (Figs. 4 and 5). For some challenging tasks like recognizing $L1_0$ LCOs in the Au-Cu with weak segregation degree and small size (below 2-nm diameter), a lower threshold, like 0.3, can ensure that atoms belonging to the hidden nanostructures are recognized as complete as possible. To further validate the reasonability of the selected threshold of 0.3, we analyzed the z-SDMs of the atoms with a threshold between 0.3 and 0.5. The double interplanar spacing still existed for this part of data like Fig. 7c, suggesting its nature of $L1_0$-typed structure.

Our previous work [25, 46] has the ability to detect LCOs, even smaller chemical short-range orders (approx. 0.5 nm in radius), focusing more on structural information. However, as highlighted in the introduction, this necessitates the transformation of 3D point cloud data into 1D signals, a computationally intensive process that is circumvented in the proposed AtomNet. While AtomNet facilitates the detection of LCOs in the Au-Cu alloy, its optimal performance lies in characterizing chemical medium-range orders, approximately 0.8 nm in radius. AtomNet exhibits less satisfactory performance in recognizing smaller chemical short-range orders, attributed to the disturbance caused by lateral atoms with lower resolution.

The current approach deals with FCC-based alloys, but it can be easily extended to other structures (BCC, HCP) without limiting the number of components. This would broaden the capability and application of AtomNet, including e.g. compositionally-complex alloys or recognizing grain boundaries in nanocrystalline materials. Moreover, better performance of AtomNet could be achieved with more realistic training datasets, which can be synthesized via advanced generation models like generative adversarial networks [74] and diffusion models [75]. Finally, the recognized accuracy of AtomNet is dependent on the data quality, like the detection efficiency and spatial resolutions. With the improvement of data quality, the boundary of AtomNet can be pushed to more complex situations, even for the detection of 1D (like dislocations) and 0D (like vacancies) nanoscale features.

## 5. Conclusions

In this work, we designed a 3D deep neural network named AtomNet to intelligently mine hidden nanoscale 3D/2D features from APT data in various FCC-based metallic materials. During training, a crucial feature updating strategy was introduced to achieve a better recognition ability. AtomNet considers both the compositional and structural information, and enables to recognize different microstructures at the singe-atom level, ranging from nanoprecipitates in the AlLiMg, LCOs in the Au-Cu, and even 2D SFs in the Co-based superalloy. AtomNet outperforms previous isosurfaces and CNN-APT methods in its ability to detect nanoprecipitates independently of the presence of crystallographic orientations and to reveal small LCOs without obvious elemental segregation. AtomNet has the ability to display unseen structures that are not present in the training data, such as SFs in the Co-based superalloy. In the near future, AtomNet will be extended to include other crystal structures (BCC, HCP) and more complex compositions, and enable the detection of grain boundary and dislocation.

**Author contributions**
Y. L. Z. W., M. S., and B. G. designed the study. J. Y. and Y. L. developed the machine learning framework. A. S. and B. G. provided APT data and made an initial analysis. Y. L. made further APT analysis. J. Y. and Y. L. drafted the original manuscript. Y. L., Z. W., B. G., and M. S. participated in major revisions. All authors contributed to the discussion of the results and commented on the manuscript. Y. L. finalized the paper.

**Declaration of Competing Interest**
The authors declare that they have no known competing financial interests or personal relationships that could have appeared to influence the work reported in this paper.

**Acknowledgments**
Yue Li acknowledges the research fellowship provided by the Alexander von Humboldt Foundation. We also acknowledge funding from the Max Planck research network on big-data-driven materials science (BiGmax) and the Open Foundation of State Key Laboratory of Powder Metallurgy at Central South University, Changsha, China. Min Song acknowledges the financial support from the Science and Technology Innovation Program of Hunan Province (No. 2022RC3035). A.S. acknowledges the financial support from Deutsche Forschungsgemeinschaft (DFG) under project A4 of the collaborative research center SFB/TR 103. The Co-based superalloys were provided by A. Bezold and S. Neumeier as part of the collaborative work pursued in the same SFB/TR 103, Project B3. Paraskevas Kontis is acknowledged for providing the Au-Cu APT data. Stefan Bauer is acknowledged for fruitful discussion.

## Data availability
The key data that support the findings are involved in this paper. Other data are available from the corresponding authors upon reasonable request.

## Code availability
The AtomNet code is available at https://github.com/bookofstrange/AtomNet.

A Case Study in Superalloys, JOM 70(9) (2018) 1736-1743.